%% file: main.tex
\documentclass{article}
\usepackage{spconf,amsmath,graphicx,hyperref}

\usepackage{multirow}
\usepackage{xcolor}
\usepackage{booktabs}

\title{Location Matters: Layer-wise Prompt Conditioning for Voice--Style Disentanglement in Text-to-Speech}

\name{David Lindevelt, Suzan Verberne, Joost Broekens}
\address{
     \textit{Leiden University}\\
     \textit{LIACS} \\
     Leiden, The Netherlands
}

\begin{document}

\maketitle

\begin{abstract}
	Independent control of voice and emotional style characteristics of speech is a
	long-standing problem in text-to-speech (TTS) systems. We investigate how the
	conditioning location of voice and style prompts across TTS decoder layers
	affects the trade-off between preservation and expressiveness for voice
	identity and emotional style.
	We introduce a layer-dependent gated fusion mechanism and systematically study
	which layers should be influenced by which prompt type.
	Our findings show an asymmetric relationship. Voice benefits from early
	conditioning for identity preservation, while style benefits from early
	conditioning for expressiveness.
	Our findings provide practical guidance that helps neural network architects
	make an appropriate trade-off for the architecture of TTS systems according to
	their use cases.
\end{abstract}


\section{Introduction}

Text-to-Speech (TTS) systems have made impressive progress in the past years.
Applications ranging from audiobook narration, podcast generation to dubbing,
advertisement voice-overs and conversational agents increasingly demand
independent control over speaker identity and expressive
style~\cite{xieControllableSpeechSynthesis2025}. For instance, in audiobook
generation, a single character voice needs to convey varying emotions across
scenes while preserving the voice identity. For conversational agents,
appropriate emotions must be independent of the speaker's voice to create
engaging experiences. Current TTS systems, however, entangle these features,
which hinders their applicability in real world scenarios.

Recent TTS systems control voice and style characteristics through
natural-language prompts, but the effect of where and how strongly these
prompts are injected into the model architecture remains underexplored. In this
paper, we propose and evaluate a method for layer-wise conditioning with
natural-language prompts to enable better control over emotional expressiveness
and speaker characteristics. Lyth et
al.~\cite{lythNaturalLanguageGuidance2024a} condition a codec-based TTS on a
single prompt covering speaker identity, style and recording conditions,
without separating those factors.
PromptTTS~\cite{guoPromptttsControllableTextToSpeech2023} defines style through
five factors and augments expert-written prompts with a language model.
InstructTTS~\cite{yangInstructTTSModellingExpressive2024} trains a prompt
encoder and applies mutual information minimization to separate style from
speaker identity. PromptTTS++~\cite{shimizuPromptTTSControllingSpeaker2024},
whose LibriTTS-P~\cite{kawamuraLibriTTSPCorpusSpeaking2024} annotations we use,
conditions on separate speaker and style prompts. All four works inject the
prompt at a single fixed point in the architecture.

A second line of work varies the granularity of conditioning. DiffStyleTTS
~\cite{liuDiffStyleTTSDiffusionbasedHierarchical2025} models prosody at two
levels with adjustable classifier-free guidance, showing that level-wise
control beats global conditioning, but does not address voice--style
disentanglement. StyleFusion-TTS ~\cite{chenStyleFusionTTSMultimodal2025} is
closest to our work: it fuses speaker and style embeddings hierarchically,
rendering speaker before style on the principle that identity is coarse
guidance and style is fine customization. That ordering is fixed by design and
never varied. We explicitly study the effect of variation.

Our contributions are therefore: \textbf{1.} We introduce a layer-dependent
gated fusion mechanism that controls the extent to which a layer is conditioned
by a prompt. \textbf{2.} We systematically investigate how conditioning
\emph{location} affects control over, and disentanglement of, voice and style.
\textbf{3.} We show that voice and style should be conditioned at different
layers for transformer-based token prediction models.

\section{Method}
We propose a conditioning paradigm called \textbf{Gated-Fusion} to control the
influence of voice and style prompts at different layers of the TTS model. As
TTS backbone, we use a transformer-based encoder-decoder architecture inspired
by Battenberg et al.~\cite{battenbergRobustUnboundedLength2025} with learnable
position encoding. The model operates on discrete
WavTokenizer~\cite{jiWavTokenizerEfficientAcoustic2025} codes. Our decoder
consists of 6 decoder blocks with 16 attention heads and hidden width 1024. The
alignment block uses a single 256-width LSTM layer and 4 heads in the
location-based attention mechanism. The encoder contains 2 residual convolution
stages that down-sample the input phoneme sequence by 2x in time, followed by 3
non-causal self-attention blocks with 8 heads and width 512.

We condition our models with natural language prompts for voice and style
separately. Each prompt is processed by a T5-large
encoder~\cite{raffel2020exploring}. Gated-Fusion processes each prompt
separately, and fuses the resulting embeddings with layer-dependent gating
weights before cross-attention, where each token has information from both
signals. We use non-learnable gating weights to isolate the effect of
conditioning location from any impact introduced by learned gates. The fusion
signal is computed as:
\begin{equation}
	\text{FS} = \sigma(\alpha_l) \cdot \text{SSP} + (1 - \sigma(\alpha_l)) \cdot \text{VSP}
\end{equation}
Where $\text{FS}$ is the fused signal, $\text{SSP}$ and $\text{VSP}$ are style and voice signals projections, $\sigma$ is a sigmoid function, and $\alpha_l$ is the gating weight for layer $\text{l}$. Style and voice signal projections are obtained by projecting style and voice signals into the same dimension as cross-attention conditioning.

We define 4 model variants for comparison. Baseline \textbf{Concat}
concatenates two prompts into a single unified text for cross-attention,
allowing the model to freely attend to any part of the combined prompt (no
gating mechanism). The remaining three use Gated-Fusion by fixing gating
weights across the six decoder layers: \textbf{Equal} sets $\alpha_l = 0$ for
all layers ($\sigma(\alpha_l) = 0.5$); \textbf{Style}-early sets $\alpha_l =
	10$ for layers 1–3 and $\alpha_l = -10$ for layers 4–6, yielding near-binary
gating ($\sigma(±10) \approx 1/0$) with style dominating early layers and voice
dominating later layers; \textbf{Voice}-early inverts this assignment.

\section{Experimental Setup}
\textbf{Data} \label{subsec:data} We use
LibriTTS-R~\cite{koizumiLibriTTSRRestoredMultiSpeaker2023} audio samples
together with LibriTTS-P~\cite{kawamuraLibriTTSPCorpusSpeaking2024} human
annotations. Note that the human annotations were created at the speaker level,
not at the individual samples. The dataset contains $515.1$ hours of speech
from $2295$ speakers, with a total of $331{,}141$ samples.

\textbf{Feature Design and Prompt Creation} \label{subsec:prompt_creation} We
describe voice identity and speaking style with four discrete features each.
\textbf{Style} features are \textit{Speed}, \textit{Loudness}, \textit{Arousal}
and \textit{Valence}. \textbf{Voice} is described with \textit{Age},
\textit{Sex}, \textit{Pitch} and \textit{Texture}. Speaking rate, loudness, and
pitch (with Penn~\cite{morrisonCrossdomainNeuralPitch2024}) are measured from
the audio; arousal and valence predicted with the Odyssey
model~\cite{goncalvesOdyssey2024Speech2024}. The other features are derived
from LibriTTS-P\cite{kawamuraLibriTTSPCorpusSpeaking2024} annotations through a
scoring function that normalizes for how many keywords each annotator uses.
Pitch is discretized within annotated sex, so the category is relative to
speakers of the same voice type rather than to an absolute threshold. Every
feature is discretized into three categories at the 25th and 75th percentiles,
except sex, which is bipolar and uses fixed thresholds. Full extraction and
scoring details and the supplementary material are in the accompanying
repository\footnote{\href{https://github.com/dcze4/ICASSP_location_matters/tree/4b10df117efc0622c808e0808296a1a00759a4fd}{Supplementary
		code repository}}.

Four style features with three values give 81 style prompts; four voice
features, with at most one texture keyword, result in 243 voice prompts. We
compare two prompting strategies. \textbf{Structured} prompts concatenate the
feature categories directly, isolating architectural effects from prompt
variation. \textbf{Template-based} prompts render the same categories through
20 style and 15 voice handwritten sentence templates, randomized per sample
during training, introducing the free-text diversity a deployed system would
face.

Pitch enters the experiment twice, but as two different quantities: the voice
prompt conditions on the speaker's absolute register (median $F_0$ over their
ten longest samples), while the style representation used for evaluation
(Section~\ref{subsec:evaluation}) measures within-utterance $F_0$ variation in
semitones relative to each utterance's own mean. Because mean-normalization
removes absolute register, the style dimensions cannot respond to the pitch
token in the voice prompt.

\textbf{Evaluation Framework} \label{subsec:evaluation} We measure control with
two metrics defined using pairwise distances in a proxy embedding space.
\textbf{Spread} is the mean pairwise distances between samples generated from
\textit{different} prompts and measures how much the model varies its output;
\textbf{drift} is the mean pairwise distance between samples with the
\textit{same} prompt and measures how much the corresponding attribute changes
when only the other prompt is varied.

Let $\mathbf{x}_{v,s}$ be a sample synthesized with voice prompt $v \in V$ and
style prompt $s \in S$. Fixing each voice and varying style,
\begin{equation}
	\bar{D}(\mathbf{e}, d) = \frac{1}{|V|} \sum_{v \in V}
	\frac{2}{|S|(|S|-1)} \sum_{i < j}
	d\!\left(\mathbf{e}(\mathbf{x}_{v,s_i}),\;
	\mathbf{e}(\mathbf{x}_{v,s_j})\right)
\end{equation}
gives Style Spread on style embeddings and Voice Drift on voice
embeddings; fixing style and varying voice gives Voice Spread and Style Drift.
A well disentangled model should have high spread on the varied factor and low drift on the fixed one.

To avoid conclusions that depend on a single embedding space, we compute every
metric in several. Voice is represented by ECAPA-TDNN, TitaNet and
WavLM~\cite{desplanquesECAPATDNNEmphasizedChannel2020,
	koluguriTitaNetNeuralModel2022, chenWavLMLargeScaleSelfSupervised2021} speaker
embeddings under cosine distance. Style is represented by seven
acoustic-prosodic features aligned with the style prompt (syllable rate, energy
mean and standard deviation, pitch standard deviation and range, arousal,
valence), with arousal and valence from either
Audeering~\cite{wagnerDawnTransformerEra2023} or
Odyssey~\cite{goncalvesOdyssey2024Speech2024}; we use Euclidean distance on
z-normalized features, since cosine distance discards magnitude differences
that are meaningful on these dimensions.

Because the same prompt units and generated audio recur across conditions, the
observations are clustered rather than independent. We therefore model each
metric with a linear mixed-effects model with a random intercept per prompt
unit and a variance component for shared generated audio, using conditioning
strategy and proxy embedding model as fixed effects. We test the
conditioning-strategy effect and its interaction with the proxy embedding model
by likelihood-ratio tests, and compare strategies with Holm-corrected pairwise
contrasts. As a robustness check we re-estimate all fixed effects with
generalized estimating equations using cluster-robust standard errors;
conclusions are unchanged (supplementary).

\textbf{Model Training and Inference Details} \label{subsec:training} All model
variants are trained on a single A100 GPU for 150k iterations. For the
generation of the speech, we use a temperature of 1.0, top-$p=0.95$ and
top-$k=200$.

\section{Results and Analysis} \label{sec:results}

\textbf{(1) Conditioning location affects voice and style control differently}
\label{subsec:condition_location} We fix input text and vary the prompts to
investigate the effects of conditioning location on the synthesized output and
report each metric averaged over the fixed factor. Table~\ref{tab:voice_emb}
shows that the Voice-early model preserves voice identity best against a
variety of style prompts, achieving the lowest drift across all three proxy
embedding spaces (ECAPA-TDNN: $.227$; TitaNet: $.470$; WavLM: $.125$). The
ranking of the models for voice drift holds across all embedding spaces. The
order shows us that voice drift falls as voice conditioning moves to earlier
layers. This suggests the model locks in speaker identity early in the forward
pass, and allows subsequent layers (where style conditioning is stronger) to
modulate prosody without disturbing identity representation. In contrast, voice
spread rankings are inconsistent across proxy spaces. This indicates that
conditioning location has limited influence on how much voice identity varies
across different voice prompts.

\input{tables/tab_mixedlm} \input{tables/tab_contrasts}

\input{tables/tab_voice_emb} \input{tables/tab_style_emb}

Table~\ref{tab:style_emb} shows that rankings are preserved across proxy spaces
for both spread and drift metrics. Style-early achieves the highest spread
(Audeering: $3.536$; Odyssey: $3.532$), while Voice-early the lowest. For style
drift, Concat is the lowest, and Equal the highest. These patterns indicate
that style spread improves when the style signal is introduced early in the
architecture, while style drift falls when style conditioning is placed later
after voice conditioning. The consistency of these rankings across proxy models
with different architectures confirms that the observed differences stem from
conditioning location rather than proxy model bias.

The mixed-effects analysis (Table~\ref{tab:mixedlm}) confirms a significant
effect of conditioning location on every metric (LRT $\chi^2(3) =
	104.8$--$411.3$, all $p < .001$). For style, the effect is strongest on spread
($\chi^2 = 411.3$) and remains large for drift ($\chi^2 = 216.5$). The random
effects show how strongly the observations are clustered: between-unit
differences account for 34--72\% of the variance and the shared generated audio
for up to 53\%, which is why the metrics are modeled with random effects rather
than as independent observations. For voice metrics, the proxy embedding model
remains the dominant fixed effect due to large scale differences between
embedding spaces; after accounting for it, voice drift retains a strong
placement effect ($\chi^2 = 366.9$) while voice spread shows the weakest of the
four ($\chi^2 = 104.8$, with all pairwise differences below $.01$). This
confirms the inconsistent rankings of voice spread in
Table~\ref{tab:voice_emb}. Post-hoc comparisons (Table~\ref{tab:contrasts})
reveal the asymmetry: the largest pairwise difference is seen in style spread
($\Delta M = -.325$) for Style-early vs.\ Voice-early, while voice drift
differences remain small ($\Delta M = -.023$) and voice spread differences are
negligible ($\Delta M = .003$). This indicates that swapping conditioning
location has large effects on style expressiveness but only a modest effect on
voice preservation.

\textbf{(2) The model needs architectural pressure to disentangle}
\label{subsec:architectural_pressure} We compare Concat and Equal as baselines
to understand whether signal separation alone affects disentanglement.
Tables~\ref{tab:voice_emb} and \ref{tab:style_emb} show that the Concat model
achieves the lowest style drift, but the highest voice drift, while the Equal
model shows the opposite pattern: highest style drift but second-lowest voice
drift. This suggests that when the model receives both signals without explicit
separation, it favors encoding style over voice, possibly because prosodic
features are more immediately useful for next-token prediction than global
identity. Separating the signals forces the model (Equal) to attend to voice
information, but at the cost of increased style drift. Post-hoc comparisons
(Table~\ref{tab:contrasts}) confirm this: the Equal vs.\ Concat difference is
largest for style drift ($\Delta M = -.246$) and style spread ($\Delta M =
	-.153$). The corresponding voice differences reach significance under the mixed
model but are far smaller ($\Delta M = .017$ and $-.003$). Separation primarily
redistributes how the model handles style; reducing voice drift requires not
just separation but early placement as shown above.

\input{figures/fig_style_spread} Template prompting tests whether the
architectural effects observed with structured prompts survive under more
realistic conditioning.

\textbf{(3) Effect of prompt diversity} \label{subsec:templates} Placement
effects shrink sharply under template prompting (Table~\ref{tab:mixedlm}). For
style, the LRT statistic drops from $411.3$ to $63.8$ for spread and from
$216.5$ to $22.6$ for drift. Prompt diversity introduces variance that
partially overrides the architectural steering observed with structured
prompts. Voice drift is the exception: its placement effect is essentially
unchanged ($366.9 \rightarrow 351.7$, a 4\% reduction), making voice identity
preservation through early conditioning the most robust of the four metrics to
prompt diversity. Post-hoc comparisons (Table~\ref{tab:contrasts}) reflect this
robustness. All six voice drift contrasts remain significant, whereas style
drift retains only two and voice spread three; the Style-early vs.\ Voice-early
voice drift difference is the largest of them ($\Delta M = -.028$). For style
spread, the surviving significant pairs cluster into two groups: Concat vs.
Style and Equal vs. Voice are not significantly different within each group,
but the groups differ from each other. Notably, Fig.~\ref{fig:style_spread}
shows that only Concat and Voice-early models show increased style spread under
template prompting, while Equal and Style-early models show reduced spread,
indicating an asymmetric effect of prompt diversity across model architectures.
Also, input text is itself a competing style signal. When prompts are fixed,
varying the text produces style distances approaching the style spread under
neutral prompts (supplementary, Section~7).

\section{Conclusions} \label{sec:conclusion}
We show that layer-dependent conditioning with natural-language prompts has an
asymmetric effect on voice and style disentanglement in TTS: voice benefits
from early placement for identity preservation, style benefits from early
placement for expressiveness. The model prioritizes style over voice when
signals are combined, requiring architectural pressure for disentanglement.

Voice identity and speaking style operate at different scopes. Voice is a
global, time-invariant property that holds across an utterance, while style
varies locally. This also explains model's default behavior; style operates at
the same granularity as the token-level prediction task, making it more
actionable than global voice identity. There is no single best strategy.
Placement should be chosen by the application priority.

\textbf{Limitations} We use an encoder-decoder transformer architecture;
findings may differ for other architectures. We use fixed gating weights in the
fusion mechanism to isolate conditioning location effects. Learnable weights
might achieve better configurations but would introduce additional noise. We
rely on (standard) proxies rather than perceptual listening tests, which limits
our ability to confirm whether the measured differences are perceptually
meaningful. Our training data is limited to LibriTTS-R, which is predominantly
neutral audiobook speech. Corpora with more diverse emotional and stylistic
content may yield different patterns.

\bibliographystyle{IEEEbib} \bibliography{bib/refs}

\end{document}

%% file: tables/tab_mixedlm.tex
\begin{table}[t]
	\centering
	\small
	\setlength{\tabcolsep}{3.5pt}
	\caption{Mixed-effects analysis. Likelihood-ratio tests for conditioning
		strategy (Model, $df=3$) and its interaction with the proxy embedding model
		(Model $\times$ PEM; $df=3$ for style, $df=6$ for voice), with variance shares
		of the random effects. All $\chi^2$ significant at $p<.001$.}
	\label{tab:mixedlm}
	\begin{tabular}{ll|rr|rrr}
		\hline
		                                                                     &              & \multicolumn{2}{c|}{LRT $\chi^2$} & \multicolumn{3}{c}{Variance share}                       \\
		\cline{3-4} \cline{5-7}
		                                                                     & Metric       & Model                             & M$\times$PEM                       & Unit & Audio & Res. \\
		\hline
		\multirow{4}{*}{\rotatebox[origin=c]{90}{\makebox[0pt]{Structured}}} & Style spread & 411.3                             & 98.7                               & .38  & .53   & .09  \\
		                                                                     & Style drift  & 216.5                             & 35.1                               & .72  & .24   & .04  \\
		                                                                     & Voice spread & 104.8                             & 104.8                              & .37  & ---   & .63  \\
		                                                                     & Voice drift  & 366.9                             & 311.6                              & .34  & .25   & .41  \\
		\hline
		\multirow{4}{*}{\rotatebox[origin=c]{90}{\makebox[0pt]{Templates}}}  & Style spread & 63.8                              & 102.3                              & .50  & .42   & .08  \\
		                                                                     & Style drift  & 22.6                              & 120.7                              & .84  & .13   & .03  \\
		                                                                     & Voice spread & 125.8                             & 124.6                              & .45  & ---   & .55  \\
		                                                                     & Voice drift  & 351.7                             & 171.6                              & .43  & .14   & .43  \\
		\hline
		\multicolumn{7}{l}{\footnotesize Random intercept per prompt unit, variance component for shared}                                                                                  \\
		\multicolumn{7}{l}{\footnotesize generated audio (unit $\times$ model). Voice spread: the audio component}                                                                         \\
		\multicolumn{7}{l}{\footnotesize was zero and is omitted. $n=1944, 648, 972, 2916$ in row order.}                                                                                  \\
	\end{tabular}
\end{table}

%% file: tables/tab_contrasts.tex
\begin{table}[t]
	\centering
	\small
	\setlength{\tabcolsep}{3.5pt}
	\caption{Pairwise model contrasts from the mixed-effects model (Holm-corrected, FWER $= 0.05$), --- $=$ insignificant.}
	\label{tab:contrasts}
	\begin{tabular}{lll|cc|cc}
		\hline
		                                                                     &         &         & \multicolumn{2}{c|}{Style} & \multicolumn{2}{c}{Voice}                                         \\
		\cline{4-7}
		                                                                     & Group 1 & Group 2 & Spread                     & Drift                     & Spread            & Drift             \\
		\hline
		\multirow{6}{*}{\rotatebox[origin=c]{90}{\makebox[0pt]{Structured}}} & Equal   & Concat  & $-.153$                    & $-.246$                   & $-.003$           & $\phantom{-}.017$ \\
		                                                                     & Equal   & Style   & $\phantom{-}.071$          & $-.114$                   & $-.009$           & $\phantom{-}.008$ \\
		                                                                     & Equal   & Voice   & $-.254$                    & $-.217$                   & $-.007$           & $-.015$           \\
		                                                                     & Concat  & Style   & $\phantom{-}.224$          & $\phantom{-}.131$         & $-.006$           & $-.009$           \\
		                                                                     & Concat  & Voice   & $-.101$                    & $\phantom{-}.029$         & $-.004$           & $-.033$           \\
		                                                                     & Style   & Voice   & $-.325$                    & $-.103$                   & $\phantom{-}.003$ & $-.023$           \\
		\hline
		\multirow{6}{*}{\rotatebox[origin=c]{90}{\makebox[0pt]{Templates}}}  & Equal   & Concat  & $\phantom{-}.073$          & $\phantom{-}.046$         & $\phantom{-}.008$ & $\phantom{-}.008$ \\
		                                                                     & Equal   & Style   & $\phantom{-}.064$          & ---                       & ---               & $\phantom{-}.020$ \\
		                                                                     & Equal   & Voice   & ---                        & $\phantom{-}.049$         & ---               & $-.008$           \\
		                                                                     & Concat  & Style   & ---                        & ---                       & $-.008$           & $\phantom{-}.013$ \\
		                                                                     & Concat  & Voice   & $-.083$                    & ---                       & $-.009$           & $-.015$           \\
		                                                                     & Style   & Voice   & $-.075$                    & ---                       & ---               & $-.028$           \\
		\hline
		\multicolumn{6}{l}{\footnotesize Values show $\Delta M$; all $p_{\text{Holm}} < .05$.}                                                                                                    \\
	\end{tabular}
\end{table}

%% file: tables/tab_voice_emb.tex
\begin{table}[t]
	\small
	\centering
	\setlength{\tabcolsep}{3pt}
	\caption{Voice Embeddings Pairwise Distance mean ± std. Drift (D.) and Spread (S.)}
	\label{tab:voice_emb}
	\begin{tabular}{|c|cc|cc|cc|}
		\hline
		\multirow{2}{*}{\textbf{Model}} & \multicolumn{2}{|c|}{\textbf{EC$^*$}} & \multicolumn{2}{|c|}{\textbf{TN$^*$}} & \multicolumn{2}{|c|}{\textbf{WL$^*$}}                                                                                        \\
		\cline{2-3} \cline{4-5} \cline{6-7}
		                                & \textbf{D. $\downarrow$}              & \textbf{S.  $\uparrow$}               & \textbf{D. $\downarrow$}              & \textbf{S. $\uparrow$}     & \textbf{D. $\downarrow$}   & \textbf{S.  $\uparrow$}    \\
		\hline
		Concat                          & .250 \tiny{±.025}                     & .353 \tiny{±.006}                     & .525 \tiny{±.039}                     & \textbf{.624} \tiny{±.020} & .146 \tiny{±.016}          & .264 \tiny{±.010}          \\
		Style                           & .250 \tiny{±.019}                     & .352 \tiny{±.006}                     & .505 \tiny{±.028}                     & .613 \tiny{±.020}          & .139 \tiny{±.018}          & .257 \tiny{±.007}          \\
		Voice                           & \textbf{.227} \tiny{±.023}            & .349 \tiny{±.005}                     & \textbf{.470} \tiny{±.046}            & .621 \tiny{±.016}          & \textbf{.125} \tiny{±.022} & .260 \tiny{±.005}          \\
		Equal                           & .242 \tiny{±.022}                     & \textbf{.365} \tiny{±.007}            & .491 \tiny{±.038}                     & .618 \tiny{±.020}          & .136 \tiny{±.016}          & \textbf{.267} \tiny{±.007} \\
		\hline
		\multicolumn{7}{l}{$^*$EC=ECAPA-TDNN, TN=TitaNet, WL=WavLM}
	\end{tabular}
\end{table}

%% file: tables/tab_style_emb.tex
\begin{table}[t]
	\caption{Style Embeddings Pairwise Distance mean ± std}
	\centering
	\small
	\begingroup
	\setlength{\tabcolsep}{3pt}
	\begin{tabular}{|c|cc|cc|}
		\hline
		\multirow{2}{*}{\textbf{Model}} & \multicolumn{2}{|c|}{\textbf{Audeering}} & \multicolumn{2}{|c|}{\textbf{Odyssey}}                  \\
		\cline{2-3} \cline{4-5}
		                                & \textbf{Drift $\downarrow$}
		                                & \textbf{Spread $\uparrow$}
		                                & \textbf{Drift $\downarrow$}              & \textbf{Spread $\uparrow$}                              \\

		\hline Concat                   & \textbf{2.897} \tiny{±.192}              & 3.300 \tiny{±.205}                     &
		\textbf{2.815} \tiny{±.185}     & 3.320 \tiny{±.182}                                                                                 \\ Style & 3.033 \tiny{±.213}
		                                & \textbf{3.536} \tiny{±.233}              & 2.941 \tiny{±.198}                     & \textbf{3.532}
		\tiny{±.230}                                                                                                                         \\ Voice        & 2.941 \tiny{±.184}                    & 3.239 \tiny{±.242} & 2.828
		\tiny{±.164}                    & 3.179 \tiny{±.218}                                                                                 \\ Equal        & 3.133 \tiny{±.176} & 3.461
		\tiny{±.239}                    & 3.070 \tiny{±.148}                       & 3.466 \tiny{±.225}                                      \\ \hline
	\end{tabular}
	\endgroup
	\label{tab:style_emb}
\end{table}

%% file: figures/fig_style_spread.tex
\begin{figure}[t]
	\centering
	\includegraphics[width=0.8\columnwidth]{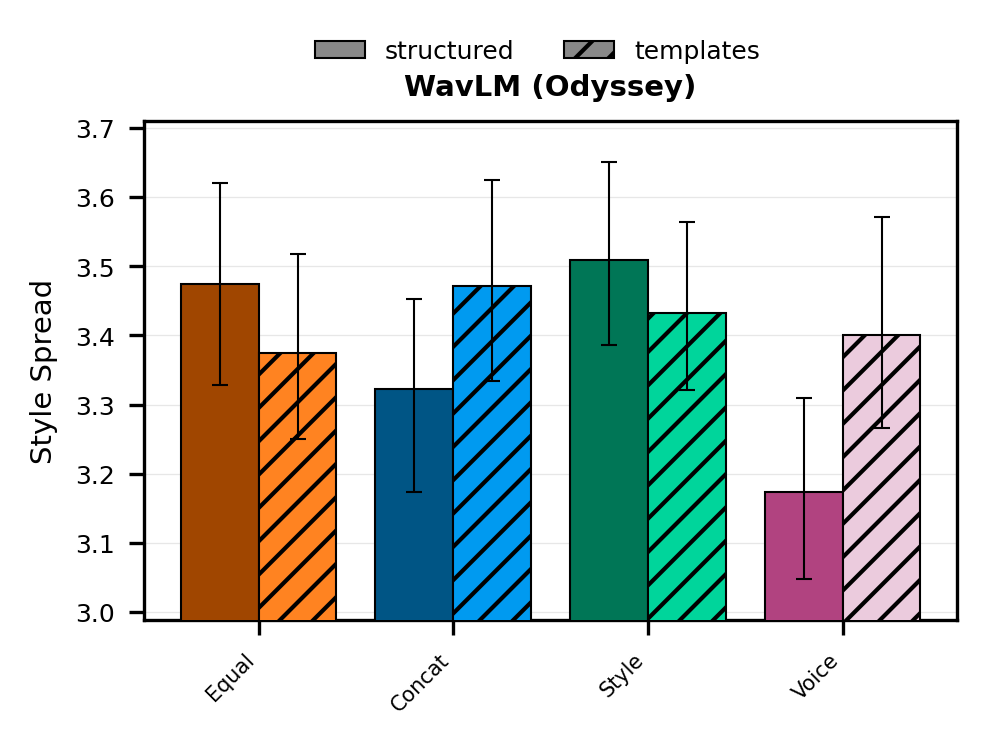}
	\caption{Style spread across conditioning strategies for structured and template prompting.}
	\label{fig:style_spread}
\end{figure}

%% file: main.bbl
\begin{thebibliography}{10}

\bibitem{xieControllableSpeechSynthesis2025}
Tianxin Xie, Yan Rong, Pengfei Zhang, Wenwu Wang, and Li~Liu,
\newblock ``Towards {{Controllable Speech Synthesis}} in the {{Era}} of {{Large Language Models}}: {{A Systematic Survey}},''
\newblock in {\em Proceedings of {{EMNLP}} 2025}, Suzhou, China, 2025, pp. 764--791.

\bibitem{lythNaturalLanguageGuidance2024a}
Dan Lyth and Simon King,
\newblock ``Natural language guidance of high-fidelity text-to-speech with synthetic annotations,'' Feb. 2024.

\bibitem{guoPromptttsControllableTextToSpeech2023}
Zhifang Guo, Yichong Leng, Yihan Wu, Sheng Zhao, and Xu~Tan,
\newblock ``Prompttts: {{Controllable Text-To-Speech With Text Descriptions}},''
\newblock in {\em {{ICASSP}}, 2023}, Rhodes Island, Greece, June 2023, pp. 1--5, IEEE.

\bibitem{yangInstructTTSModellingExpressive2024}
Dongchao Yang, Songxiang Liu, Rongjie Huang, Chao Weng, and Helen Meng,
\newblock ``{{InstructTTS}}: {{Modelling Expressive TTS}} in {{Discrete Latent Space With Natural Language Style Prompt}},''
\newblock {\em IEEE/ACM Transactions on Audio, Speech, and Language Processing}, vol. 32, pp. 2913--2925, 2024.

\bibitem{shimizuPromptTTSControllingSpeaker2024}
Reo Shimizu, Ryuichi Yamamoto, Masaya Kawamura, Yuma Shirahata, et~al.,
\newblock ``{{PromptTTS}}++: {{Controlling Speaker Identity}} in {{Prompt-Based Text-To-Speech Using Natural Language Descriptions}},''
\newblock in {\em {{ICASSP}}, 2024}, Seoul, Korea, Republic of, Apr. 2024, pp. 12672--12676, IEEE.

\bibitem{kawamuraLibriTTSPCorpusSpeaking2024}
Masaya Kawamura, Ryuichi Yamamoto, Yuma Shirahata, Takuya Hasumi, and Kentaro Tachibana,
\newblock ``{{LibriTTS-P}}: {{A Corpus}} with {{Speaking Style}} and {{Speaker Identity Prompts}} for {{Text-to-Speech}} and {{Style Captioning}},''
\newblock in {\em {{INTERSPEECH}} 2024}. Sept. 2024, pp. 1850--1854, ISCA.

\bibitem{liuDiffStyleTTSDiffusionbasedHierarchical2025}
Jiaxuan Liu, Zhaoci Liu, Yajun Hu, Yingying Gao, Shilei Zhang, and Zhenhua Ling,
\newblock ``{{DiffStyleTTS}}: {{Diffusion-based Hierarchical Prosody Modeling}} for {{Text-to-Speech}} with {{Diverse}} and {{Controllable Styles}},''
\newblock in {\em Proc. {{COLING}}}, Jan. 2025, pp. 5265--5272.

\bibitem{chenStyleFusionTTSMultimodal2025}
Zhiyong Chen, Xinnuo Li, Zhiqi Ai, and Shugong Xu,
\newblock ``{{StyleFusion TTS}}: {{Multimodal Style-Control}} and {{Enhanced Feature Fusion}} for {{Zero-Shot Text-to-Speech Synthesis}},''
\newblock in {\em Pattern {{Recognition}} and {{Computer Vision}}}, vol. 15041, pp. 263--277. Springer Nature Singapore, Singapore, 2025.

\bibitem{battenbergRobustUnboundedLength2025}
Eric Battenberg, {\relax RJ}~{Skerry-Ryan}, Daisy Stanton, Soroosh Mariooryad, et~al.,
\newblock ``Robust and {{Unbounded Length Generalization}} in {{Autoregressive Transformer-Based Text-to-Speech}},''
\newblock in {\em Proc. NAACL-HLT, 2025}, Albuquerque, New Mexico, Apr. 2025, pp. 11789--11806.

\bibitem{jiWavTokenizerEfficientAcoustic2025}
Shengpeng Ji, Ziyue Jiang, Wen Wang, Yifu Chen, Minghui Fang, Jialong Zuo, et~al.,
\newblock ``{{WavTokenizer}}: An {{Efficient Acoustic Discrete Codec Tokenizer}} for {{Audio Language Modeling}},''
\newblock in {\em The {{Thirteenth International Conference}} on {{Learning Representations}}}, 2025.

\bibitem{raffel2020exploring}
Colin Raffel, Noam Shazeer, Adam Roberts, Katherine Lee, Sharan Narang, Michael Matena, et~al.,
\newblock ``Exploring the limits of transfer learning with a unified text-to-text transformer,''
\newblock {\em Journal of Machine Learning Research}, vol. 21, no. 140, pp. 1--67, 2020.

\bibitem{koizumiLibriTTSRRestoredMultiSpeaker2023}
Yuma Koizumi, Heiga Zen, Shigeki Karita, Yifan Ding, Kohei Yatabe, Nobuyuki Morioka, et~al.,
\newblock ``{{LibriTTS-R}}: {{A Restored Multi-Speaker Text-to-Speech Corpus}},''
\newblock in {\em {{INTERSPEECH}} 2023}. Aug. 2023, pp. 5496--5500, ISCA.

\bibitem{morrisonCrossdomainNeuralPitch2024}
Max Morrison, Caedon Hsieh, Nathan Pruyne, and Bryan Pardo,
\newblock ``Cross-domain {{Neural Pitch}} and {{Periodicity Estimation}},'' Aug. 2024.

\bibitem{goncalvesOdyssey2024Speech2024}
Lucas Goncalves, Ali~N. Salman, Abinay~Reddy Naini, Laureano {Moro-Vel{\'a}zquez}, Thomas Thebaud, Paola Garcia, et~al.,
\newblock ``Odyssey 2024 - {{Speech Emotion Recognition Challenge}}: {{Dataset}}, {{Baseline Framework}}, and {{Results}},''
\newblock in {\em The {{Speaker}} and {{Language Recognition Workshop}} ({{Odyssey}} 2024)}. June 2024, pp. 247--254, ISCA.

\bibitem{desplanquesECAPATDNNEmphasizedChannel2020}
Brecht Desplanques, Jenthe Thienpondt, and Kris Demuynck,
\newblock ``{{ECAPA-TDNN}}: {{Emphasized Channel Attention}}, {{Propagation}} and {{Aggregation}} in {{TDNN Based Speaker Verification}},''
\newblock in {\em {{INTERSPEECH}} 2020}. Oct. 2020, pp. 3830--3834, ISCA.

\bibitem{koluguriTitaNetNeuralModel2022}
Nithin~Rao Koluguri, Taejin Park, and Boris Ginsburg,
\newblock ``{{TitaNet}}: {{Neural Model}} for {{Speaker Representation}} with {{1D Depth-Wise Separable Convolutions}} and {{Global Context}},''
\newblock in {\em {{ICASSP}}, 2022}, May 2022, pp. 8102--8106.

\bibitem{chenWavLMLargeScaleSelfSupervised2021}
Sanyuan Chen, Chengyi Wang, Zhengyang Chen, Yu~Wu, Shujie Liu, Zhuo Chen, et~al.,
\newblock ``{{WavLM}}: {{Large-Scale Self-Supervised Pre-Training}} for {{Full Stack Speech Processing}},''
\newblock {\em IEEE Journal on Selected Topics in Signal Processing}, vol. 16, no. 6, pp. 1505--1518, 2021.

\bibitem{wagnerDawnTransformerEra2023}
Johannes Wagner, Andreas Triantafyllopoulos, Hagen Wierstorf, Maximilian Schmitt, Felix Burkhardt, et~al.,
\newblock ``Dawn of the {{Transformer Era}} in {{Speech Emotion Recognition}}: {{Closing}} the {{Valence Gap}},''
\newblock {\em IEEE Transactions on Pattern Analysis and Machine Intelligence}, vol. 45, no. 9, pp. 10745--10759, Sept. 2023.

\end{thebibliography}
